**Detailed and high-throughput measurement of composition dependence of magnetoresistance and spin-transfer torque using a composition-gradient film: application to $Co_xFe_{1-x}$ ($0 \leq x \leq 1$) system**

Vineet Barwal[1], Hirofumi Suto[1]*, Tomohiro Taniguchi[2]*, and Yuya Sakuraba[1]

[1] Research Center for Magnetic and Spintronic Materials, National Institute for Materials Science (NIMS), Tsukuba, 305-0047, Japan

[2] National Institute of Advanced Industrial Science and Technology (AIST), Research Center for Emerging Computing Technologies, Tsukuba, 305-8568, Japan

*SUTO.Hirofumi@nims.go.jp (H. S.) *tomohiro-taniguchi@aist.go.jp (T. T.)

We develop a high-throughput method for measuring the composition dependence of magnetoresistance (MR) and spin-transfer-torque (STT) effects in current-perpendicular-to-plane giant magnetoresistance (CPP-GMR) devices and report its application to the CoFe system. The method is based on the use of composition-gradient films deposited by combinatorial sputtering. This structure allows the fabrication of devices with different compositions on a single substrate, drastically enhancing the throughput in investigating composition dependence. We fabricated CPP-GMR devices on a single GMR film consisting of a $Co_xFe_{1-x}$ ($0 \leq x \leq 1$) composition-gradient layer, a Cu spacer layer, and a NiFe layer. The MR ratio obtained from resistance-field measurements exhibited the maximum in the broad Co concentration range of $0.3 \leq x \leq 0.65$. In addition, the STT efficiency was estimated from the current to induce magnetization reversal of the NiFe layer by spin injection from the $Co_xFe_{1-x}$ layer. The STT efficiency was also the highest around the same Co concentration range as for the MR ratio, and this correlation was theoretically explained by the change in the spin polarization of the $Co_xFe_{1-x}$ layer. The results revealed the $Co_xFe_{1-x}$ composition range suitable for spintronic applications, demonstrating the advantages of the developed method.

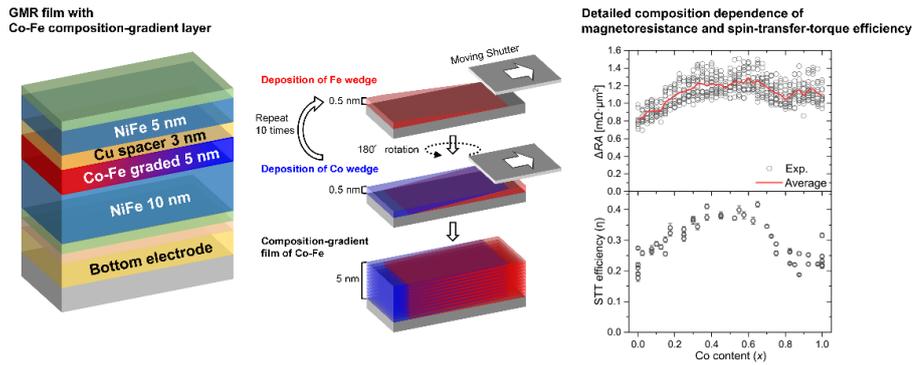

## 1. Introduction

The exploration of new avenues in material research is driven by developing efficient methodologies that overcome conventional limitations in throughput and fine data granularity. Combinatorial sputtering, a high-throughput materials synthesis technique that creates a library sample containing variations in the material parameters, is one such tool and has recently been focused in the field of magnetism and spintronics [1]–[4]. It can efficiently generate large datasets in combination with local and automated measurements, which are becoming increasingly important from the perspective of recent data-driven approaches such as machine learning, as datasets are the basis for predicting new promising materials. This approach has been applied to study spin-orbit torque switching, magnetic anisotropy, and saturation magnetization, as these parameters are the key to improving the performance in applications such as magnetic recording, magnetic memory, and permanent magnets [5]–[8].

This study focuses on magnetoresistance (MR) and spin-transfer torque (STT) effects in MR devices. These effects are of technological importance and serve as the operating principle in magnetoresistive random access memory [9], hard-disk-drive read heads [10], computing devices [11], spin-torque oscillators [12], and other spintronic devices. The effect of composition has been one of the main topics to improve the device performance, and material systems such as CoFe [13]–[19], CoMnSi [20]–[22], CoMnFeSi [23], CoMnFeGe [24], CoFeGaGe [25], and MnGa [26] have been investigated. However, a problem with high-throughput experiments is that multiple MR stacks must be deposited for each composition point to be studied when the experiments are performed on a uniform composition sample. One deposition can take up to one day if it uses in-situ annealing. In addition, MR measurement requires microfabrication of the MR stacks into pillar devices with a typical diameter of 100 nm, as will be explained later in the experimental part of this paper. This process also takes several days. Subsequently, MR measurements can take several hours if the statistical data points are corrected by manual probing. Therefore, revealing detailed composition dependence from many samples requires a lot of time and effort, and methods to overcome this limitation can significantly advance material research for spintronic applications.

In the present work, we developed a high-throughput method for measuring the detailed composition dependence of MR and STT effects in current-perpendicular-to-plane giant magnetoresistance (CPP-GMR) devices. Our method is based on the use of a GMR stack containing a composition-gradient layer fabricated by a cluster-type combinatorial sputtering system with ten cathodes. Owing to this sample structure, the CPP-GMR devices with composition variation at fine intervals are prepared on a single substrate by a single set of microfabrication. The MR measurements on these devices are then performed by an automatic probing system. The method drastically enhances the throughput in the investigation of composition dependence.

We applied the developed method to the CoFe system, which has been a widely used material in spintronics. Its high spin polarization is desirable for achieving large MR and STT [27], [28]. With a low damping parameter, it offers the advantage of reduced operational currents in devices[29]. Furthermore, its high saturation magnetization is particularly beneficial for assisted writing in magnetic recording applications [30]–[33].The composition dependence of CoFe has been investigated to improve the MR and STT property [23], which makes CoFe the ideal test system for the developed method. Although the previous experimental reports agreed in that the MR ratio is higher around the $Co_{50}Fe_{50}$ composition than at the Co or Fe sides, no comprehensive and detailed composition dependence has been reported, as samples with only a few different compositions were prepared and compared in each study. [13]–[19]

We fabricated CPP-GMR devices on a GMR stack with a $Co_xFe_{1-x}$ ($0 \leq x \leq 1$) composition-gradient layer. Measurements on these devices on the single substrates provided composition dependence over the whole range at a fine interval of $x = 0.025$. The MR ratio exhibited the maximum in the broad Co concentration range of $0.3 \leq x \leq 0.65$. In addition to the MR ratio, STT efficiency was evaluated by using magnetization reversal measurements, which we recently proposed [34]. The STT efficiency exhibited a similar trend to the MR ratio, and their correlation was theoretically explained by the change in the spin polarization of $Co_xFe_{1-x}$. The results efficiently revealed the $Co_xFe_{1-x}$ composition range suitable for spintronic applications, demonstrating the usefulness of the developed method.

Here, we comment on previous studies on the characterization of magnetic thin films using combinatorial sputtering. H. Masuda et al. investigated spin-Hall effect in Cu-Ir binary alloys by spin Peltier imaging and found the optimum Ir concentration around 25 at.% for enhanced SHE [3]. T. Scheike et al. reported MR property of magnetic tunnel junctions containing thickness gradient of MgO barrier [4]. However, study on compositional dependent properties in multilayer CPP-GMR stacks via combinatorial technique has not been reported. Technically, the proposed method is more relevant to the operation of the actual devices fabricated from the multilayer stacks as compared to the studies on single layer thin films generally used as combinatorially sputtered samples. In addition, because of the smaller MR ratio of GMR devices and larger distributions due to nanofabrication, statistical analysis based on a larger number of devices is necessary to discuss the trend.

## 2. Sample fabrication and measurement method

The GMR stack shown in Fig. 1(a) was deposited using a combinatorial sputtering system onto a thermally oxidized Si substrate with a planarized Cu/Ta bottom electrode. The spin-injection layer consisted of a 10 nm $Ni_{0.8}Fe_{0.2}$ layer and a 5 nm $Co_xFe_{1-x}$ ($0 \leq x \leq 1$) composition gradient layer. Figure 1(b) shows the procedure to make the $Co_xFe_{1-x}$ layer. Wedge-shaped Co and Fe layers with linear thickness variation from 0 to 0.5 nm along the $X$-direction were deposited by moving a shutter during deposition. A pair of Co and Fe layers forms a flat 0.5-nm-thick layer by reversing the wedge directions between the Co and Fe layers. The pairs were deposited repeatedly ten times to form a 5-nm-thick $Co_xFe_{1-x}$ layer. Figure 1(c) shows a photo image of the sample (2 × 1 cm) after device fabrication. The sample size is 20 and 10 mm along the $X$- and $Y$-directions, respectively, and the composition gradient existed with a width of 10 mm in the center of the $X$-axis. The additional NiFe layer in the spin-injection layer was employed because of the requirements from the magnetization reversal measurements, as explained later. The spacer layer was composed of Cu, and the free layer was composed of $Ni_{0.8}Fe_{0.2}$. The composition gradient in this study was one-dimensional, as it was suitable for the binary system of CoFe. For a ternary system, a composition gradient can be made two-dimensional by depositing a wedge-shaped layer in three directions by 120°.

The GMR stack was patterned into pillars using electron beam lithography and Ar ion milling. The pillars had circular and elliptical shapes with designed dimensions of 80 × 80 nm, 140 × 70 nm, 100 × 100 nm, and 200 × 100 nm. After patterning, the pillars were passivated with a $SiO_2$ layer, and a Au top electrode was deposited. The devices were located in a rectangular grid along the $X$-direction (composition-gradient direction) and $Y$-direction. 5 devices with each size, 20 devices in total, were fabricated along the $Y$-direction, where the composition was the same. Because of the small pillar sizes, the composition was considered uniform inside the pillars, and the data from these devices represented the property of the corresponding composition. The distance between the neighboring devices in the X-direction was 250 μm, meaning that the devices with 40 different compositions are prepared at intervals of $x = 0.025$.

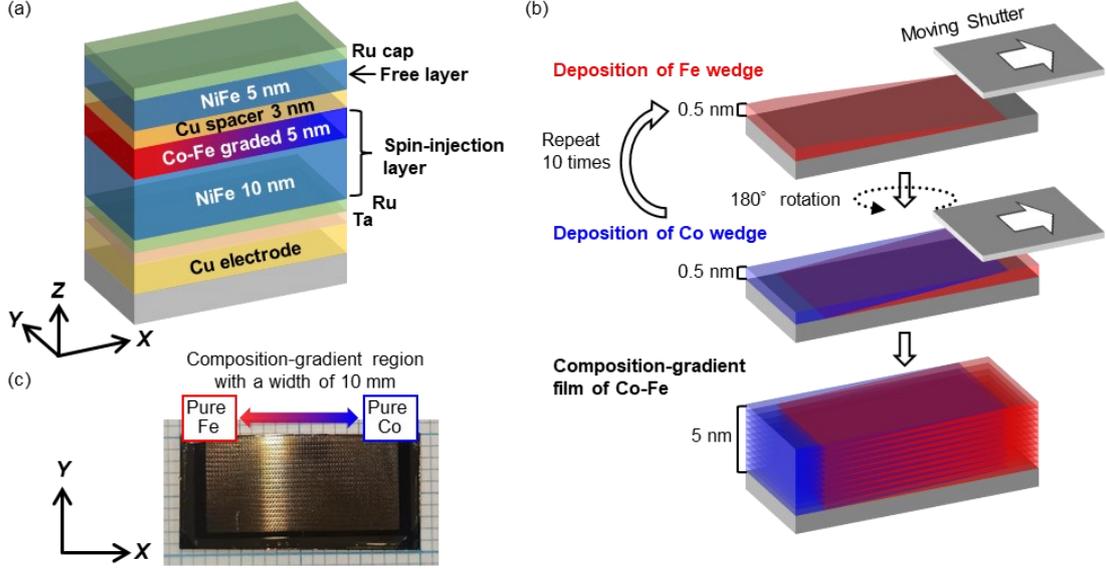

FIG. 1. (a) Configuration of the GMR stack containing a composition-gradient $Co_xFe_{1-x}$ layer. (b) Procedure to make composition-gradient $Co_xFe_{1-x}$ layer. (c) Photo image of the sample.

Resistance versus in-plane-magnetic field (R-H) measurements were carried out on the devices of all four sizes by using an automatic prober system. An in-plane magnetic field in the range of ± 90 mT was applied to the devices, and the device resistance was measured by the four-probe method.

The STT efficiency was estimated by inducing reversal of the free layer magnetization by the spin injection from the spin-injection layer, using the method we recently proposed and demonstrated [34]. In this magnetization reversal measurements, circular pillars with 80 × 80 nm diameter were used, whose device area was $10.58 \times 10^{-3}$ μm² as estimated by scanning electron microscopy. The magnetization directions of both magnetic layers were first aligned to the perpendicular direction by applying a sufficient perpendicular magnetic field ($H_z$), and the resistance versus bias voltage (R-$V_b$) measurements were carried out. By applying a sufficient $V_b$ the free layer magnetization was reversed against $H_z$ by STT, which was detected as a change of $R$ due to the MR effect. In this experiment, the NiFe layer in the spin-injection layer increased the magnetic volume to stabilize the spin-injection layer magnetization to the spin injection from the free layer. The $R$-$V_b$ curves were measured at $\mu_0 H_z$ values from 1.5 T to 3 T in steps of 0.1 T.

In the magnetization reversal process, the magnetization dynamics for the unit vector of the free layer magnetization, $\mathbf{m}_{FL}$, is described by the following Landau-Lifshitz-Gilbert (LLG) equation containing the STT term:

$$\frac{d\mathbf{m}_{FL}}{dt} = -\gamma\mu_0 \mathbf{m}_{FL} \times \mathbf{H}_{eff} + \alpha \mathbf{m}_{FL} \times \frac{d\mathbf{m}_{FL}}{dt} - \frac{\hbar \gamma J}{2|e|dM_s^{FL}} \eta \mathbf{m}_{FL} \times (\mathbf{m}_{FL} \times \mathbf{m}_{SIL}). \qquad (1)$$

The first and second terms on the right side are the precession and damping terms, respectively, where $\gamma$, $\mathbf{H}_{eff}$, and $\alpha$ are the gyromagnetic ratio, effective field, and damping constant. $\mathbf{H}_{eff}$ includes $H_z$, the demagnetizing field, the anisotropy field, and the dipolar field from the spin-injection layer, respectively. The third term represents the STT term, where $\hbar$ is Planck's constant, $J$ is the current density, $e$ is the elementary charge, $d$ is the free layer thickness, $M_s^{FL}$ is the free layer saturation magnetization, and $\mathbf{m}_{SIL}$ is the unit vector of the spin-injection layer magnetization. The STT efficiency, $\eta$, is a dimensionless quantity that represents the relation of the STT amplitude to the spin polarization of the magnetic material and to the angle between the magnetizations of the free and spin-injection layers. We consider the following condition. The spin-injection layer magnetization is in the $+z$ direction because of $H_z$, and the free layer magnetization rotates near the equator because of the precession as the damping term balances with the STT term. Under this condition, $\mathbf{H}_{eff}$ and $\mathbf{m}_{SIL}$ effectively contain only a $z$-direction component. Then, the critical current density $J_c$ that satisfies the balance between the damping and STT terms is expressed as:

$$J_\text{c} = \mu_0 \frac{2|e|}{\hbar\eta} \alpha M_s^{\text{FL}} d H_\text{eff}. \qquad (2)$$

This equation indicates that $J_\text{c}$ is proportional to $H_\text{eff}$ and thus has a linear dependence on $H_z$. Therefore, $\eta$ can be estimated from the slope of the linear relation. The advantage of measuring the slope is that it excludes the effect of the dipolar field from the spin-injection layer and the demagnetizing field of the free layer, which are included in $H_\text{eff}$. Because the accurate estimation of these two fields is difficult as the former depends on the $M_s$ of the spin-injection layer and changes with the $Co_xFe_{1-x}$ composition, and the latter depends on the pillar size. Considering the slope can exclude their effects because they only affect the intercept.

## 3. Results and discussion

We first present the *R-H* measurement results. Figure 2(a) shows the *R-H* curves from the three devices with different compositions of Fe, $Co_{0.525}Fe_{0.475}$, and Co, respectively. Around zero field, all the devices exhibited the antiparallel configuration of the free and spin-injection layers due to the dipolar interaction between them, resulting in a high *R* state. By applying *H*, magnetizations moved to the field direction, thereby decreasing the *R*, and a low *R* state corresponding to the parallel configuration was observed at $\mu_0 H \sim \pm 30$ mT.

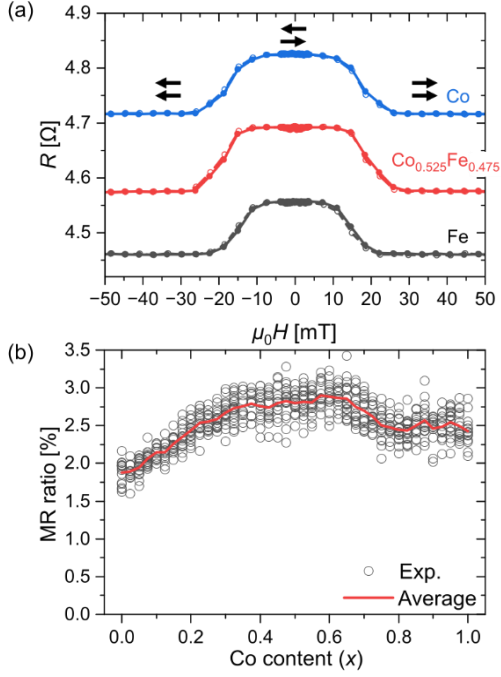

FIG. 2. (a) in-plane *R-H* curves obtained from the circular devices with compositions of Fe, Co$_{0.525}$Fe$_{0.45}$, and Co. The designed diameter is 80 nm. Solid and dashed lines represent the upward and downward field sweeps, respectively, and the curves from the two sweep directions almost overlap. Data are offset for clarity, and arrows on top represents the magnetization configurations. (b) MR ratio as a function of Co content. Symbols represent the experimental data, and a line represents the average.

Figure 2(b) shows the composition dependence of the MR ratio calculated from the *R-H* measurements. Among the 20 devices fabricated along the *Y*-direction having the same composition, those showing a deviation over 20% from the average were considered defective, and their data were excluded. The result revealed the complete composition dependence as follows. On the Fe side, the MR ratio was the lowest and around 1.7%. The MR ratio increased with *x* and reached the maximum of approximately 2.8% for $0.3 \leq x \leq 0.65$. Then the MR ratio dropped for $0.65 \leq x \leq 0.8$, and the MR ratio was almost constant at around 2.25% for $0.8 \leq x \leq 1$. The result was consistent with the previous studies referred to in the introduction part, which reported the trend at several composition points [13], [15], [18].

We next present the results of the magnetization reversal measurements. Figures 3(a)-3(c) show the $R$-$V_b$ curves at different $H_z$ values obtained from the three devices with compositions of Fe, $Co_{0.525}Fe_{0.475}$, and Co, respectively. These results were obtained from the same sample in the $R$-$H$ measurements shown in Fig. 2(a). The $R$-$V_b$ curves exhibited an overall parabolic increase by the temperature increase due to Joule heating. The additional $R$ increase appears on applying sufficiently large negative $V_b$, which corresponds to the reversal of free layer magnetization due to the spin injection. As observed in the $R$-$V_b$ curves, the $V_b$ amplitude required to induce the magnetization reversal increases with $H_z$.

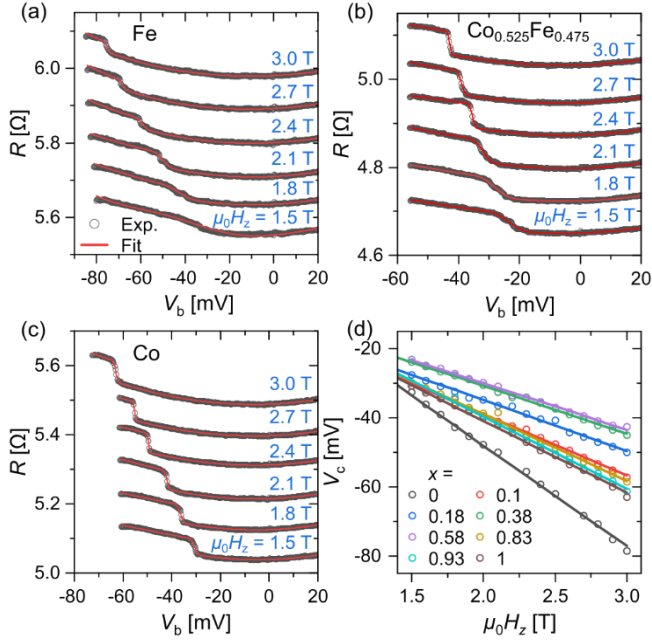

FIG. 3. $R$-$V_b$ curves obtained from the devices with compositions of (a) Fe, (b) $Co_{0.525}Fe_{0.475}$, and (c) Co, respectively. Data obtained in different $H_z$ values are shown with offset for clarity. Open circle symbols represent the experimental data, and lines represent fitting using Eq. (3). (d) $V_c$ versus $H_z$ obtained from CPP-GMR devices with different compositions. Symbols represent the experimental data, and lines represent linear fitting.

The $R$-$V_b$ curves were fitted phenomenologically using the following equation:

$$R = f(V_b) + \frac{\Delta R}{2}\left(1 + \text{erfc}\left(\frac{V_b - V_c}{V_{\text{width}}}\right)\right), \tag{3}$$

where $f$ is a second-order polynomial function representing the $R$ change due to the temperature change, $\Delta R$ represents the amount of the $R$ change due to the magnetization reversal, erfc is the error function, $V_c$ corresponds to the $V_b$ value at the center of the $R$ change, and $V_{\text{width}}$ represents the $V_b$ width of the $R$ change covering approximately 85% of the $R$ change.

Figure 3(d) summarizes the $H_z$ dependence of $V_c$ for several $Co_xFe_{1-x}$ compositions. The linear relation between $V_c$ and $H_z$ was clearly observed, which is consistent with the theory of magnetization reversal mentioned above. The slope representing the STT efficiency changed with the composition. The STT efficiency was deduced from the slope using Eq. 2. In this calculation, α was set to 0.008 as reported in Ref. [35], and $\mu_0 M_s$ of NiFe free layer was taken to be 1 T. Figures 4(a) and 4(b) show the composition dependence of a change in the resistance-area product ($\Delta RA$) obtained from the $R$-$H$ measurements and $\eta$ obtained from the magnetization reversal measurements. $\Delta RA$ was calculated assuming the typical pillar size for all the sample. Here, we show $\Delta RA$ instead of MR ratio because $\Delta RA$ is comparable to the results in the previous studies, which used the samples with an antiferromagnetic pinning layer [15], [16]. The $\Delta RA$ results are also discussed later using a theoretical model. The composition dependence of $\Delta RA$ was basically the same as that of MR ratio in Fig. 2(b), with a slightly larger scattering due to the distribution in the pillar size. The trend of MR property agrees well with that of $\eta$ in terms of the increase for $0 \leq x \leq 0.3$, maximum for $0.3 \leq x \leq 0.65$, decrease for $0.65 \leq x \leq 0.8$, and plateau for $0.8 \leq x \leq 1$. This coincidence suggests that the composition dependence of the MR property and STT efficiency originated from the change in the spin polarization of $Co_xFe_{1-x}$ in the spin injection layer.

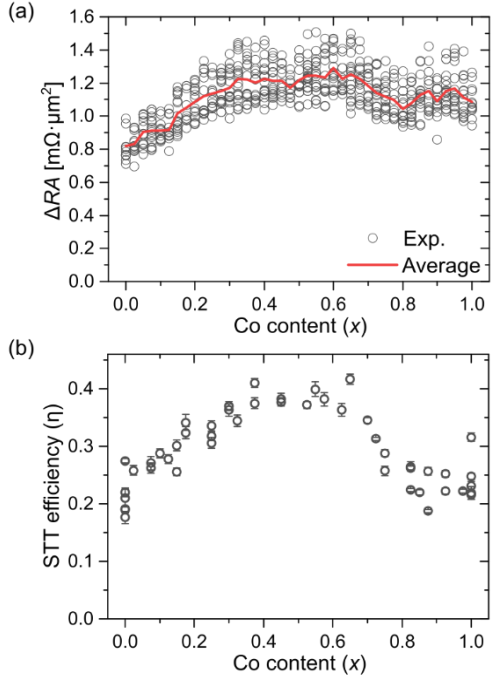

FIG. 4. (a) ΔRA as a function Co content. Symbols represent the experimental data, and a line represents the average. (b) STT efficiency parameter as a function Co content.

Although ΔRA and STT efficiency showed the similar trend, there were several differences, as follows. The rate of change was higher for STT efficiency, that is, in comparison between Fe ($x = 0$) and the maximum range for $0.3 \leq x \leq 0.65$, ΔRA increased by approximately 50 % and the STT efficiency increased by approximately. 90 %. The ΔRA at Fe was lower than that at Co, while the STT efficiency was similar between the two sides. These differences are attributed to the different mechanisms in MR and STT effects. The MR effect originates from the electrons spin parallel to the magnetization, while the STT originates from the spin perpendicular to the magnetization. Although they both depends on spin polarization, STT additionally involves spin mixing conductance. Therefore, it is crucial to evaluate these effects individually using the structure close to the actual devices for accurate prediction of their performance.

To examine the change in Δ*RA* and STT efficiency, we conducted theoretical calculations (see the supplementary material for the details.) Figures 5(a) and 5(b) show calculated Δ*RA* and $\eta$ with respect to the bulk spin polarization of the spin-injection layer ($\beta_{SIL}$). The interfacial resistance and its spin asymmetry, which also contribute to spin polarization, were fixed for simplicity. This assumption was consistent with the previous studies, which reported similar interfacial spin asymmetry in $Co_xFe_{1-x}$/Cu, at several composition points [18], [36]. The calculation results indicated that Δ*RA* and $\eta$ show similar dependence on $\beta_{SIL}$ and qualitatively explain the experimental results that Δ*RA* and $\eta$ exhibited similar composition dependence. Note that the calculation parameters were taken from the low-temperature experiments because those at room temperature were not reported. Therefore, the bulk spin polarization of $Co_xFe_{1-x}$ cannot be estimated by comparing the experimental and calculation results. The calculated Δ*RA* agreed with that reported in the low-temperature experiment [18].

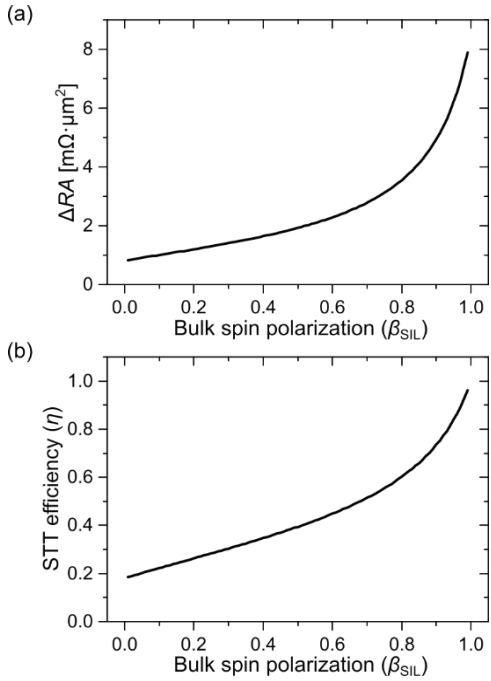

FIG. 5. Calculated (a) Δ*RA* and (b) STT efficiency parameter as a function bulk spin polarization of the spin-injection layer.

## 4. Conclusion

In summary, we have developed a high-throughput method to measure MR and STT effects in CPP-GMR devices and demonstrated its usefulness in the CoFe system, which currently plays a central role in spintronic materials. Using a GMR stack containing a composition-gradient $Co_xFe_{1-x}$ ($0 \leq x \leq 1$) layer, devices with 40 different compositions were prepared on a single substrate. The results provide a guideline for selecting the CoFe composition beneficial for spintronic applications. The conventional method of fabricating and measuring multiple samples for each composition would take 200 days to cover such a large number of compositions, and our method achieved a 40-fold increase in the throughput, demonstrating the ability of the developed method to accelerate material research in spintronic applications. The method is also applicable to other spintronic materials, such as Heusler alloys, where the composition significantly affects the material properties. In fact, we have already started applying the method to Heusler systems, and we saw a clear composition dependence in the preliminary results.

Funding

This work was supported by the Advanced Storage Research Consortium (ASRC), JSPS KAKENHI Grant No. 21K20434, JST CREST Grant No. JPMJCR21O1, and MEXT Initiative to Establish Next-generation Novel Integrated Circuits Centers (X-NICS) Grant No. JPJ011438.

Data availability statement

The data that support the findings of this study are available from the corresponding author upon reasonable request.